\documentclass{article}

\usepackage{arxiv}

\usepackage[utf8]{inputenc} 
\usepackage[T1]{fontenc}    
\usepackage[colorlinks,bookmarksopen,bookmarksnumbered,citecolor=red,urlcolor=red]{hyperref} 
\usepackage{moreverb,url}   
\usepackage{booktabs}       
\usepackage{amsfonts}       
\usepackage{amsmath}        
\usepackage{nicefrac}       
\usepackage{microtype}      
\usepackage{tikz}           

\title{Understanding performance variability in standard and pipelined parallel Krylov solvers}

\author{
  Hannah Morgan \\
  Divsion of Mathematics and Computer Science \\
  Argonne National Laboratory \\
  Lemont, IL 60439 \\
  \texttt{hmorgan@anl.gov} \\
\And
  Patrick Sanan \\
  Institut f\"ur Geophysik\\
  ETH Z\"urich\\
  8092 Z\"urich, CH \\
  \texttt{patrick.sanan@erdw.ethz.ch} \\
\And
  Matthew G.~Knepley \\
  Department of Computer Science and Engineering\\
  University at Buffalo\\
  Buffalo, NY 14260 \\
  \texttt{knepley@buffalo.edu} \\
\And
  Richard Tran~Mills \\
  Divsion of Mathematics and Computer Science \\
  Argonne National Laboratory \\
  Lemont, IL 60439 \\
  \texttt{rtmills@anl.gov} \\
}

\begin{document}
\maketitle

\begin{abstract}
In this work, we collect data from runs of {\color{blue} Krylov subspace methods and pipelined Krylov algorithms} in an effort to understand and model the impact of machine noise and other sources of variability on performance.
 We find large {\color{blue} variability of Krylov iterations between compute nodes} for standard methods that is reduced in {\color{blue} pipelined algorithms},
  directly supporting conjecture, as well as large variation between statistical distributions of runtimes across iterations.
  Based on these results, we improve upon a previously introduced nondeterministic performance model by allowing iterations to fluctuate over time.
  We present our data from runs of various {\color{blue} Krylov algorithms} across multiple platforms as well as our updated non-stationary model that provides good agreement with observations. {\color{blue} We also} suggest how it can be used as a predictive tool.
\end{abstract}

\section{Introduction}

Coherent performance models are important to describe and evaluate algorithms on modern architectures.
With good models, HPC users can pick the appropriate method and algorithmic parameters for their application to run in a complicated computing environment.
Futhermore, models are important in evaluating HPC systems and moving forward with new architecture and algorithm design.

The increasing complexity of computing environments has introduced many sources of run-to-run variability at different levels of HPC systems.
For example, an operating system can interrupt processors at any time for its activities, causing detours in computations and
degrading performance (\cite{hoefler2010characterizing, FerreiraBridgesBrightwell08}).
Inter-job contention for shared resources such as network bandwidth, routers, and links is another source of variability that
can affect application performance (\cite{parker2017early, chunduri2017run}).
Developing coherent performance models in the presence of variability is difficult
(\cite{beckman2006influence, HoeflerLumsdaineRehm07}), particularly between systems.

In this work, we advocate for casting performance models in a nondeterministic
paradigm that more accurately reflects algorithm execution in noisy HPC environments.
Employing stochastic performance models for algorithms that run on HPC systems could reflect a more realistic computing
scenario since many detours and sources of variability are unpredictable.
Some work has been done using stochastic models to predict performance in HPC environments.
For example, \cite{agarwal2005impact}  studied the impact of different noise distributions on a single collective operation and \cite{seelam2010extreme}
modeled operating system ``jitter" as random variables in order to compute computational slowdown in the presence of noise.

We are interested in modeling standard {\color{blue} Krylov subspace methods} (\cite{saad96iterative,VanDerVorst2003}) and pipelined variants often used in large-scale
simulations to solve sparse systems of equations.
Highly synchronous applications such as those that use {\color{blue} Krylov subspace methods} are vulnerable to performance degradation induced
by variability at different machine levels.
Because each iteration is performed in lockstep, an operating system interrupt (or any other source of slowdown) that delays
a single processor can cause others to be delayed as well.
{\color{blue}
The cost of communication can vary considerably between iterations and runs on modern machines due to network
contention or other factors. This} can significantly degrade method performance because of the global reduction
operations used to compute vector norms and inner products.
Motivated by mitigating latency associated with global communication, algebraically equivalent {\color{blue} pipelined
algorithms of Krylov subspace methods} have been introduced
(\cite{Chronopoulos_Gear_1989, GhyselsAshbyMeerbergenVanroose2013, GhyselsVanroose2014, StrzodkaGoddeke06,
Sturler_Vorst_1995, JacquesNicolasVollaire12}).
{\color{blue} Pipelined algorithms} rearrange computations so that it is possible to hide some of the cost of global
communication with local computation, at the cost of {\color{blue} increased computation, storage, and degraded numerical stability.}

We characterize the performance of {\color{blue} standard Krylov subspace methods and pipelined algorithms} using experimental data to refine the performance
model introduced (see \cite{morgan2016krylov}), which used random variables to model iteration times.
Such models and data are useful because pipelined {\color{blue} Krylov algorithms} tend to become more useful in situations where performance
data is more difficult to directly gather: with large node counts on congested supercomputers with unpredictable local processing times.
We intentionally maintain a data-driven approach, focusing on a simple stochastic model that nevertheless captures a key property of pipelined {\color{blue} Krylov algorithms in the context of increasingly-complex  computing systems.} This is to be strongly contrasted with traditional performance modeling which emphasizes describing best-case scenarios (\cite{GroppLusk1999}), attainable speedup, and detailed measurements of presumed-to-be-relevant performance characteristics.

Data is found to directly support the conjecture that {\color{blue} pipelined algorithms} effectively smooth out variability which in and of itself can give speedup over
standard {\color{blue} Krylov subspace methods}.
This demonstrates that in addition to helping to address the scalability bottleneck associated with global reductions, pipelined {\color{blue} Krylov
algorithms} can, by relaxing data dependencies, offer increased performance even when reductions are modeled as simple {\color{blue} barriers.
This} points to these methods being useful in a much wider range of applications, such as those with smaller problem sizes but more
variable local times per iteration, or on systems with extremely fast {\color{blue} networks that are currently} far away from a reduction latency scaling bottleneck.

In this paper, we begin by reviewing the performance models introduced in \cite{morgan2016krylov}
for {\color{blue} Krylov subspace methods and pipelined algorithms} in the presence of noise. Next we examine experimental results
from runs of {\color{blue} these algorithms} and refine the performance models based on
insights from the data. Specifically, we intend to show the need for non-stationary models.
{\color{blue} Throughout this work, we use the term ``non-stationary'' to mean a process (in our case, iterations of a Krylov algorithm) whose probability distribution changes with shifts in time.}
We employ our stochastic performance models and show that the updated model is in close
agreement with reality.  Then we present more experimental
data and suggest ways to perform a priori performance estimates.  Finally, we conclude our work.

\section{Stochastic model review} \label{sec:model-review}

A {\color{blue} standard Krylov method} is modeled as a set of $P$ processes repeatedly performing local computation and global communication. Because of the synchronizations in each of {\color{blue} iteration the} total time for each iteration is the time given by the slowest processor. Let $p$ index the $P$ processors and $k$ index the $K$ iterations. Then, the total time $T$ for a Krylov method is modeled as
\begin{equation}
T = \sum_k \max_p T^k_p,
\end{equation}
where $T^k_p$ is the time for iteration $k$ on process $p$. Since this work can be interrupted by operating system noise or other interference, the iteration times $T^k_p$ might fluctuate over steps and processors. To model this nondeterministic behavior, we let the iterates be random variables and ask for the expected total runtime
\begin{equation}
E[T] = \sum_k E[ \max_p \mathcal{T}^k_p]. \label{eq:krylov-expression}
\end{equation}
If the iterates $\mathcal{T}^k_p$ are identical and independent of processor ($p$) and stationary in time ($k$), then we can compute the expected runtime of a Krylov method using the integral expression
\begin{equation}
E[T] =  K P \int ^{\infty}_{-\infty} x F(x)^{P-1} f(x) dx \label{eq:krylov-model},
\end{equation}
where the random variables $\mathcal{T}^k_p$ are drawn from a distribution with probability density function (pdf) $f(x)$ and cumulative distribution function (cdf) $F(x)$.

Similarly, we model a pipelined Krylov {\color{blue} algorithm} as a set of $P$ communicating processors. {\color{blue} Pipelined algorithms} employ split-phase, non-blocking collectives that do not require global synchronizations. Rather, the collective operation is first initialized, say with  ${\texttt{MPI\_Iallreduce()}}$, and later finalized with ${\texttt{MPI\_Wait()}}$ so that useful work can continue between initialization and finalization.
Provided no processor falls too far behind, the total time for a {\color{blue} pipelined algorithm} $T'$ is the time for the slowest processor to perform all the given work
\begin{equation}
T' = \max_p \sum_k T^k_p.
\end{equation}
Because of detours, we do not expect the iteration times to be {\color{blue} deterministic.} Again, we ask for the expected total run time
\begin{equation}
E[T'] = E\left[\max_p \sum_k \mathcal{T}^k_p\right], \label{eq:pipeline-expression}
\end{equation}
letting the iterates be random variables drawn from a distribution.
If the iterates $\mathcal{T}^k_p$ are identically distributed with finite mean and variance, independent of process ($p$) and stationary in time ($k$), then, in the limit of large $K$, $\tfrac{1}{K}\sum_k\mathcal{T}^k_p \rightarrow \mu$ where $\mu$ is the mean of the underlying iteration time distribution with pdf $f(x)$ and cdf $F(x)$. So we have an expression to compute the expected total time for a {\color{blue} pipelined algorithm}
\begin{equation}
E[T'] \rightarrow K\mu. \label{eq:pipeline-model}
\end{equation}

For a more thorough explanation of the arguments presented in this section, see \cite{morgan2016krylov}.
In the next section we present fine-grained data collected from runs of {\color{blue} Krylov methods and pipelined algorithms.} This gives us insight into how detours affect these simulations, which assumptions above are realistic, and how the models might be refined.

\section{Experimental results} \label{sec:experimental-results}

{\color{blue} Here we present the results from experiments conducted on the Cray XC40, Theta, at the Argonne Leadership Computing Facility. We choose to run two programs with contrasting characteristics. While this is a limited representation the results are informative.}
Theta compute nodes each contain a single Xeon Phi Knights Landing (KNL) CPU (\cite{sodani2015knights})
 with 64 cores, and are connected by a Cray Aries interconnect in a dragonfly topology (\cite{alverson2012cray}).
We use the Portable, Extensible Toolkit for Scientific Computation (PETSc) (\cite{petsc-user-ref, petsc-web-page})
 version 3.10 to simulate the solution to problems with varying characteristics such as sparsity pattern.
One MPI rank per core is used throughout and we boot the nodes in quadrant clustering mode and flat memory mode, which exposes the on-die, high-bandwidth Multi-Channel DRAM (MCDRAM) as a user-addressable memory, rather than treating it as an L3 cache.
We collect iteration times $\mathcal{T}^k_p$ by inserting calls to ${\texttt{MPI\_Wtime()}}$ at the beginning and end of each Krylov cycle in PETSc's scalable linear equation solvers (KSP) component's GMRES (generalized minimal residual method) and PGMRES (pipelined GMRES) {\color{blue} algorithms} (\cite{GhyselsAshbyMeerbergenVanroose2013}).
Since changes and upgrades to HPC machines can affect performance, we note that experiments presented in this paper were performed in autumn 2018 with Cray-MPICH 7.7.3 {\color{blue} and built with the Cray C compiler wrapper. }


\subsection{One-dimensional Laplacian} \label{sec:ex23}
\begin{figure}[b]
\centering
\includegraphics[scale=0.5]{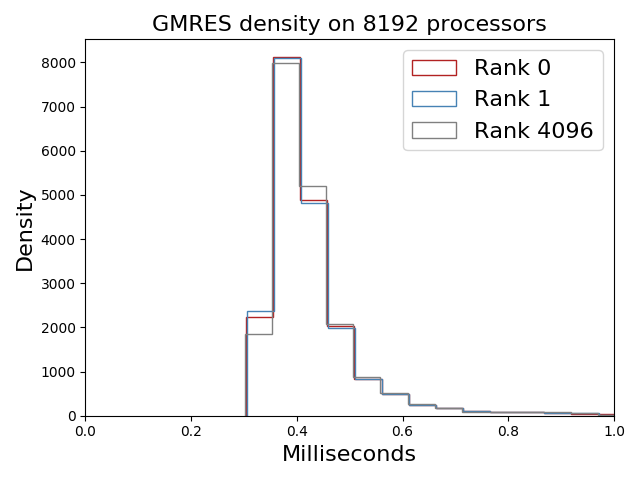}
\includegraphics[scale=0.5]{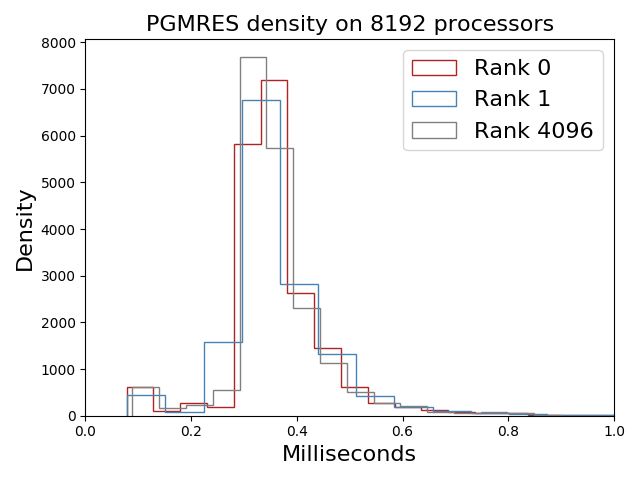}
\caption{Distribution of iterates $\mathcal{T}^k_p$ for fixed processors.} \label{fig:ex23-identical}
\end{figure}
PETSc KSP tutorial ex23\footnote{\texttt{/src/ksp/ksp/examples/tutorials/ex23.c} in PETSc 3.10} is a one-dimensional finite-difference discretization of the Laplacian, resulting in a simple tridiagonal system. A simple matrix will subtract the effects of irregular communication in sparse matrix-vector products {\color{blue} by removing the need for all but nearest neighbor communication}. We show the results from a simulation with a more complicated sparsity pattern in the next section.
We run ex23 with GMRES and PGMRES using 8192 MPI processes with $10^6$ unknowns and a Jacobi preconditioner.
Throughout these simulations we force 5000 {\color{blue} Krylov iterations}.
We choose to run our experiments with few degrees of freedom per processor since {\color{blue} these schemes} are often used in this regime and with a large number of iterations to amass statistics.

\begin{figure}[t]
 \centering
   \begin{tikzpicture}
     \node at (0, 0) {\includegraphics[scale=0.5]{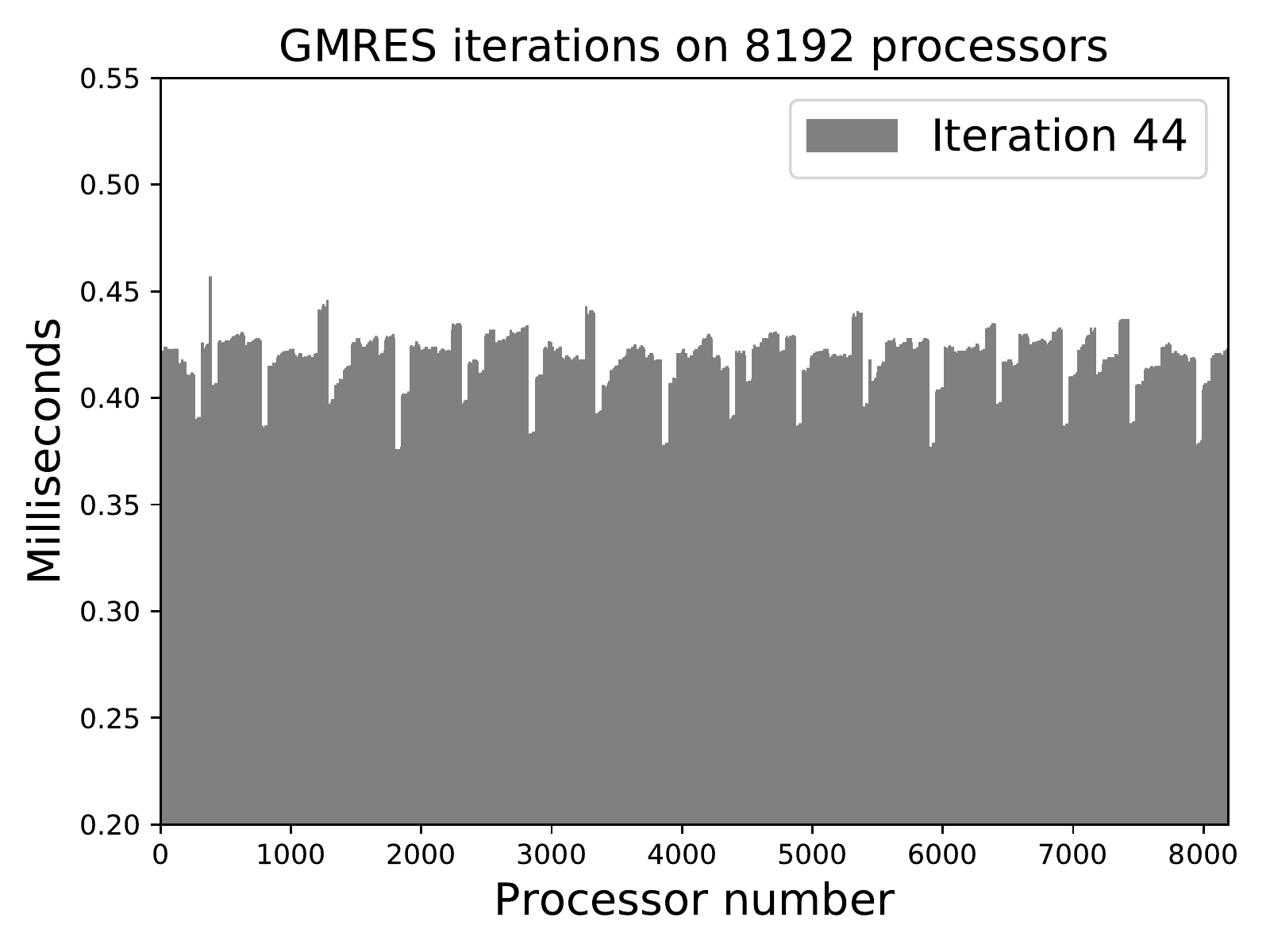}};
     \node at (-1.45, -1) {\includegraphics[width=3cm]{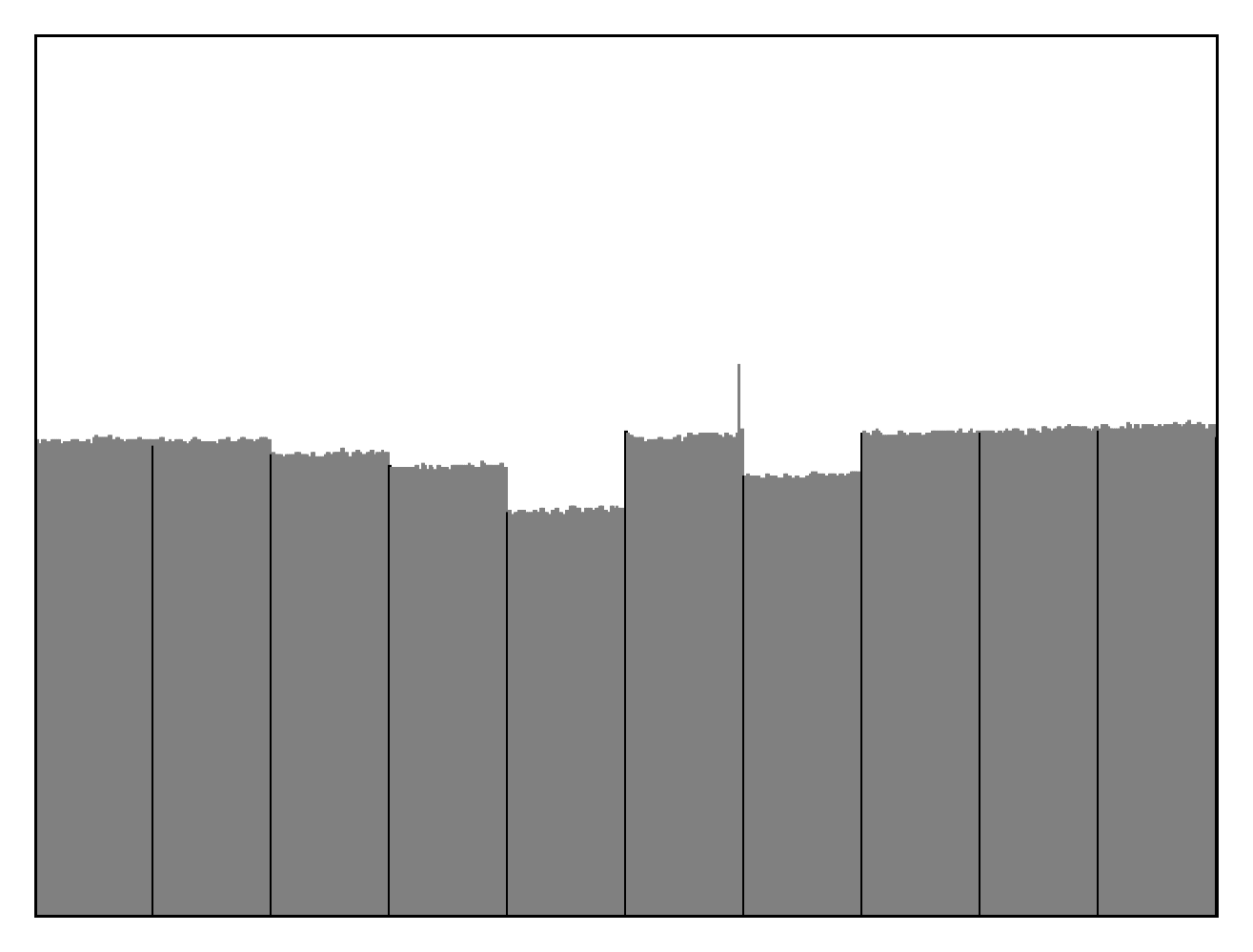}};
  \end{tikzpicture}
     \begin{tikzpicture}
     \node at (0, 0) {\includegraphics[scale=0.5]{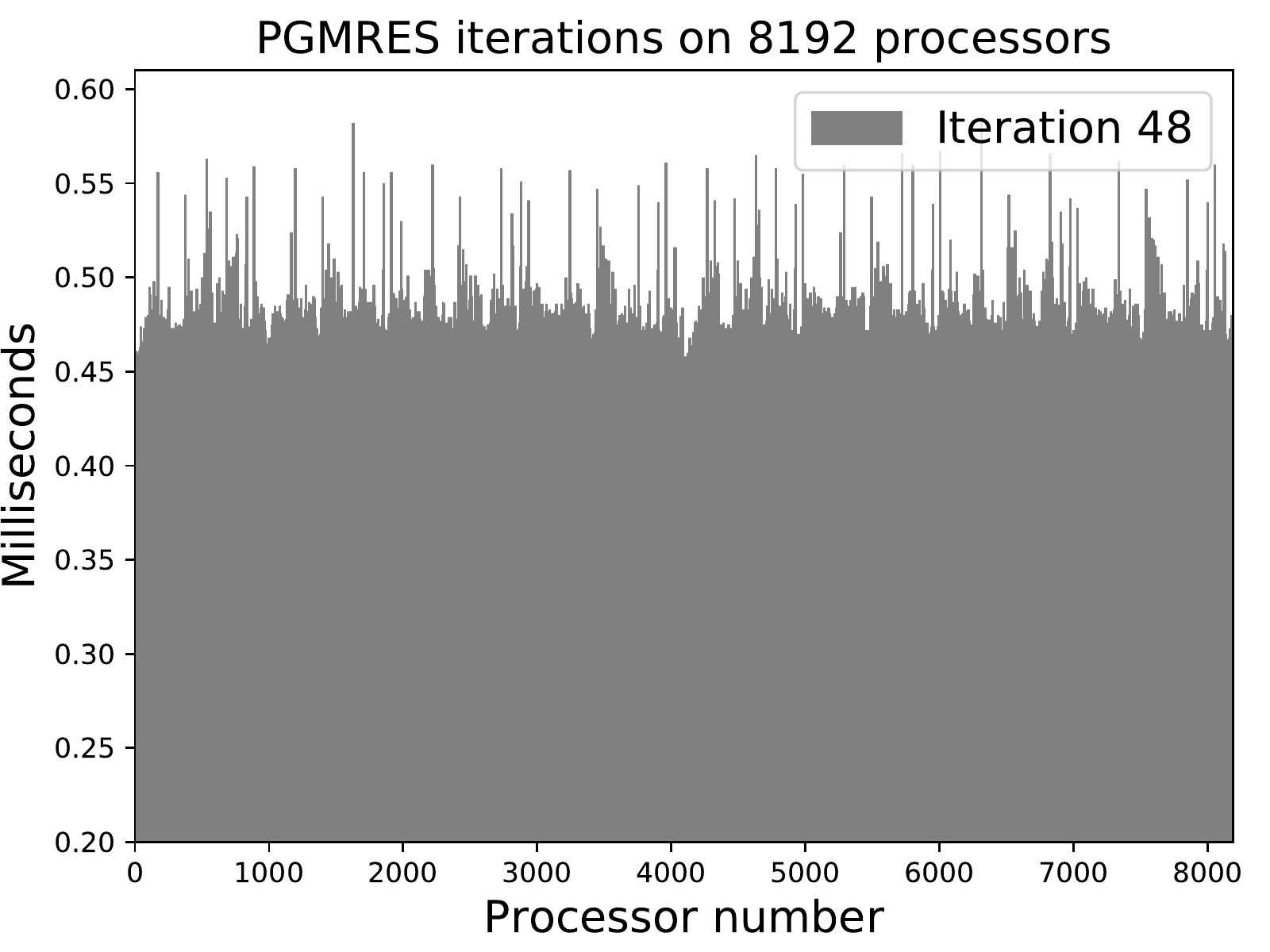}};
     \node at (-1.6, -1.1) {\includegraphics[width=3cm]{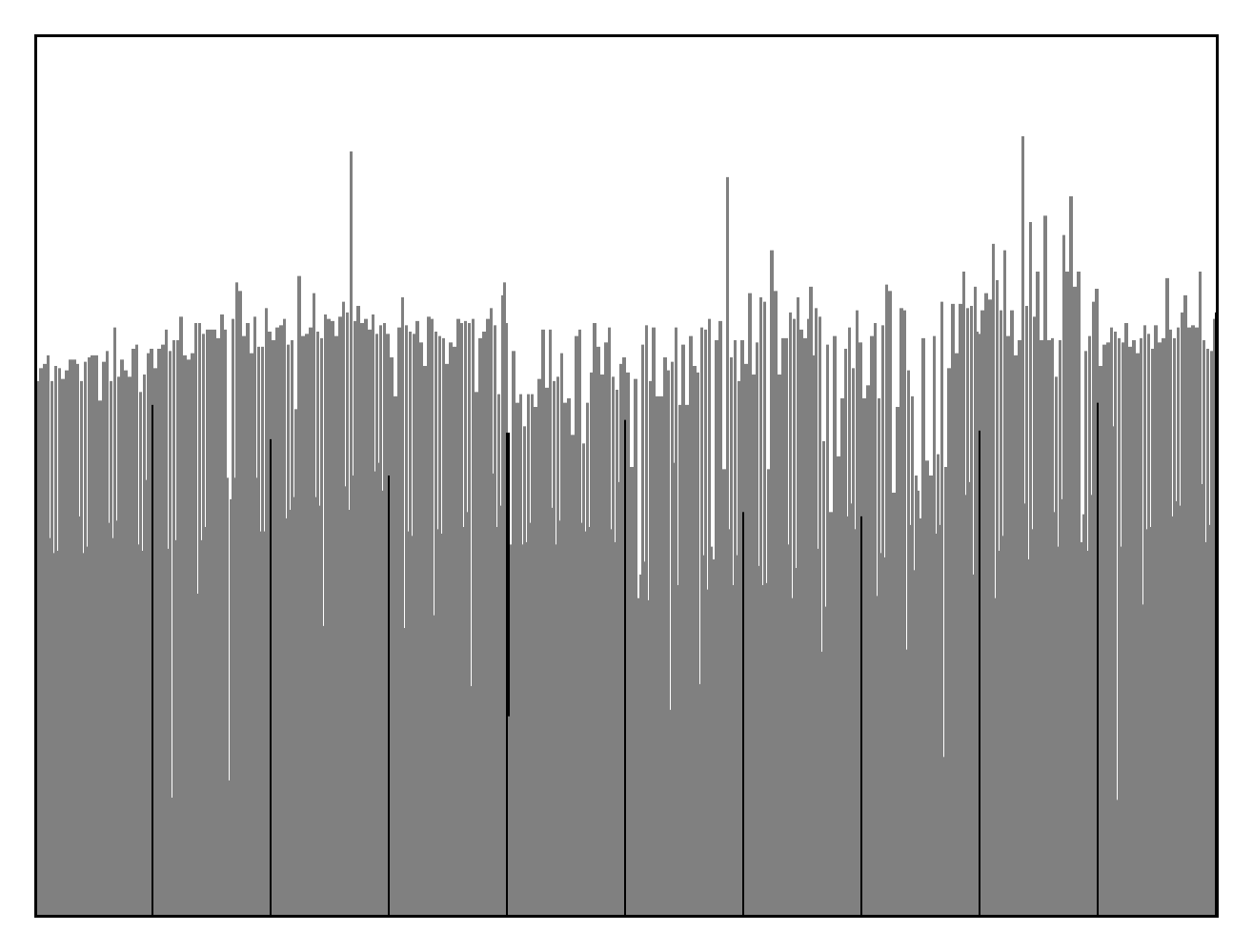}};
  \end{tikzpicture}
\caption{Time for iterations $k=44$ (GMRES) and $k=48$ (PGMRES) on process $p$ with inset showing the first 10 nodes. }
 \label{fig:ex23-independent}
\end{figure}

The stochastic model expressions \eqref{eq:krylov-model} and \eqref{eq:pipeline-model} place assumptions on the iterates $\mathcal{T}^k_p$. We check them here to examine the actual performance of the algorithms and to justify use of the models.
In Figure \ref{fig:ex23-identical} we graph the distribution of the iterates for select processors identified by their MPI ranks. The distributions look similar between ranks, implying that the iterations might be identical with respect to process. We make use of the two sample Kolmogorov-Smirnov test to check whether two samples (e.g. iterations from rank 0 and rank 1) are drawn from the same underlying distribution. The Kolmogorov-Smirnov test calculates the distance between the empirical distributions of two samples with empirical distribution functions $F_1$ of size $n$ and $F_2$ of size $m$ with
\begin{equation}
D = \sup_x |F_1(x) - F_2(x)|.
\end{equation}
We reject the hypothesis that the samples come from the same distribution with significance level $\alpha$ if
\begin{equation}
D > c(\alpha)\sqrt{\frac{n + m}{nm}}
\end{equation}
where $c(\alpha)$ is a tabulated value.
For GMRES, $n = m = 5000$ and $n = m = 5334$ for PGMRES. We use significance level $\alpha = 0.05$ and SciPy's ${\texttt{stat.ks\_2samp}}$ function to calculate the test statistic $D$.
We find that we do not reject that GMRES ranks 0 and 1 come from the same distribution with significance level $\alpha = 0.05$ since $D = 0.002 < 0.024$, the threshold. For PGMRES ranks 0 and 1, $D = 0.088 > 0.023$, so we reject the null hypothesis.
On rank 1, we reject none of the $P-1$ pairs of GMRES ranks and reject $54.5\%$ of PGMRES pairs.

Because it is inefficient to orthogonalize new Krylov vectors against a large basis, GMRES methods are typically implemented to restart after $R$ iterations. When a {\color{blue} Krylov subspace method} restarts, the basis for the Krylov {\color{blue} subspace} is discarded and the current solution is used as the initial guess for the next cycle.
PETSc implements a pipelined GMRES {\color{blue} algorithm} given by \cite{GhyselsAshbyMeerbergenVanroose2013} which adds two iterations in each cycle to "fill the pipeline."  Since the PETSc default restart value is $30$, 5334 iterations of PGMRES are actually employed instead of 5000. Some of these are very quick and easily identifiable as the bump of very short iterations in Figure \ref{fig:ex23-identical} {\color{blue} (bottom)}. Still, we compare 30 GMRES iterations to 32 of PGMRES since they compute the same number of intermediate solutions which are, moreover, the same in exact arithmetic.

Figure \ref{fig:ex23-independent} shows the time for iterations $k=44$ (GMRES) and $k=48$ (PGMRES), halfway through the Krylov cycle described {\color{blue} before}, on process $p$ in order of MPI rank.
{\color{blue} We include a inset} graph that shows the iterations on the first ten nodes, delineated by black lines.
GMRES iterates are nearly constant on a node; each KNL node of 64 processes is clearly visible in Figure \ref{fig:ex23-independent} {\color{blue} (top, inset)} so that the iterations are not truly independent in $p$ as we assumed above. The variance of iteration times on node 0 (with MPI ranks 0-63) is $6.3\cdot10^{-13} \text{s}^2$.
This could be due to the synchronizations in each iteration of {\color{blue} GMRES that force} processes on a node to perform in lockstep.
There is also some periodic behavior over all the nodes.
Conversely, PGMRES {\color{blue} iterations contain} more variability on a node ($1.4\cdot10^{-10} \text{s}^2$ on the node with MPI ranks 0-63, three orders of magnitude greater than GMRES), but less between them.

\begin{figure}[t]
\centering
\includegraphics[scale=0.5]{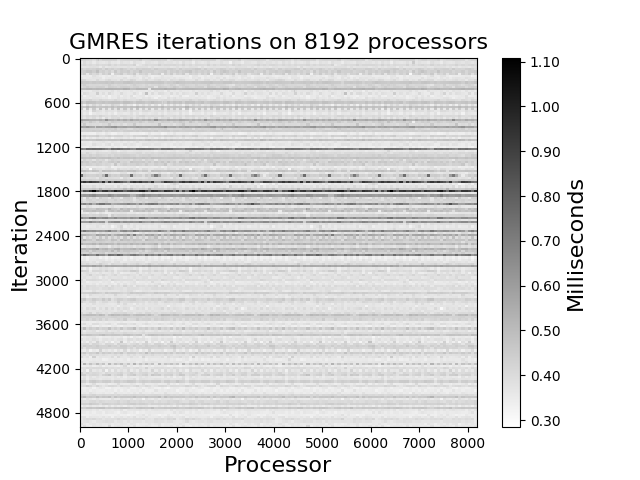}
\includegraphics[scale=0.5]{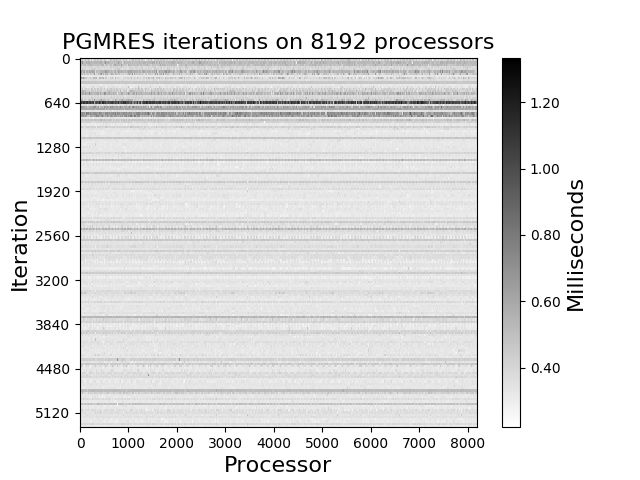}
\caption{2-color colormap of iterates $\mathcal{T}_p^k$.} \label{fig:ex23-stationary}
\end{figure}

Figure \ref{fig:ex23-stationary} is a 2-color colormap of the iterates $\mathcal{T}_p^k$ for all iterations and all processors.
Graphed are the iterations that perform the ``average" amount of work during each GMRES cycle (GMRES iteration $14 \bmod 30$ and PGMRES $16 \bmod 32$) so that each iteration shown performs the same computation.
Horizontal lines are highly visible, implying that processors perform similarly within a given iteration.
The difference between different iterations is also {\color{blue} clear; iterate timings fluctuate} in time without an obvious pattern.
{\color{blue}In the absence of an algorithmic reason that some iterations should take longer than others, after taking restarts into account, these differences could be explained by long operating system interruptions, thermal throttling of the CPU, or other external causes.
Furthermore, some groups of iterations (GMRES iterations $k \approx 1700 - 2300$ or PGMRES $k \approx 640 - 740$) overall take longer than other iterations.

In the last section we made the} simplifying assumption that the iterates were stationary in time so that the sum in \eqref{eq:krylov-expression} could become a product in \eqref{eq:krylov-model}. We account for this non-stationary {\color{blue} behavior in the next section.}

\subsection{Three-dimensional non-Newtonian fluid}

\begin{figure*}[t]
\centering
\includegraphics[scale=0.5]{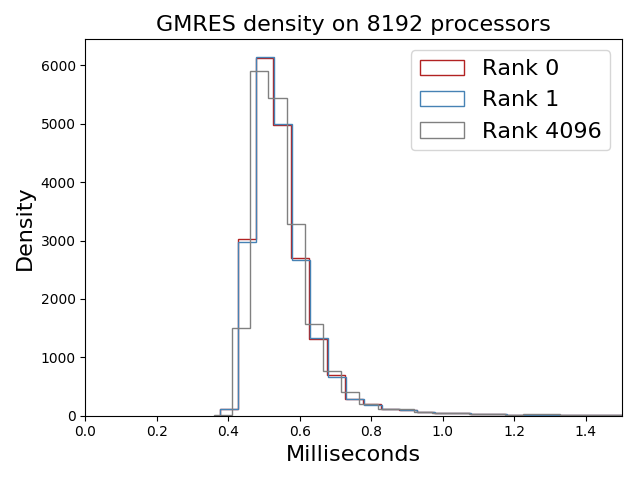}
\includegraphics[scale=0.5]{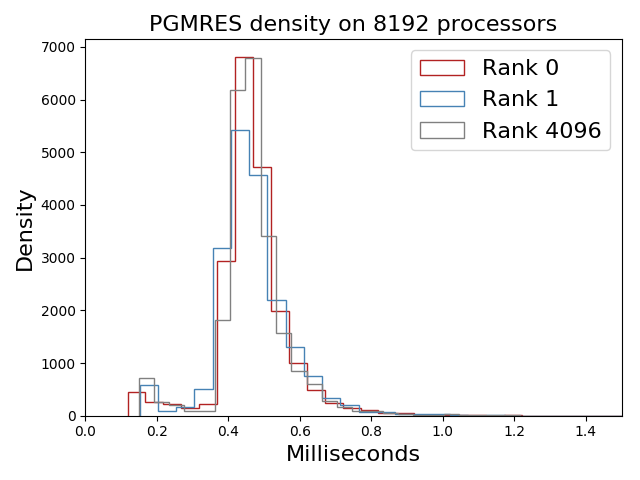}
\includegraphics[scale=0.5]{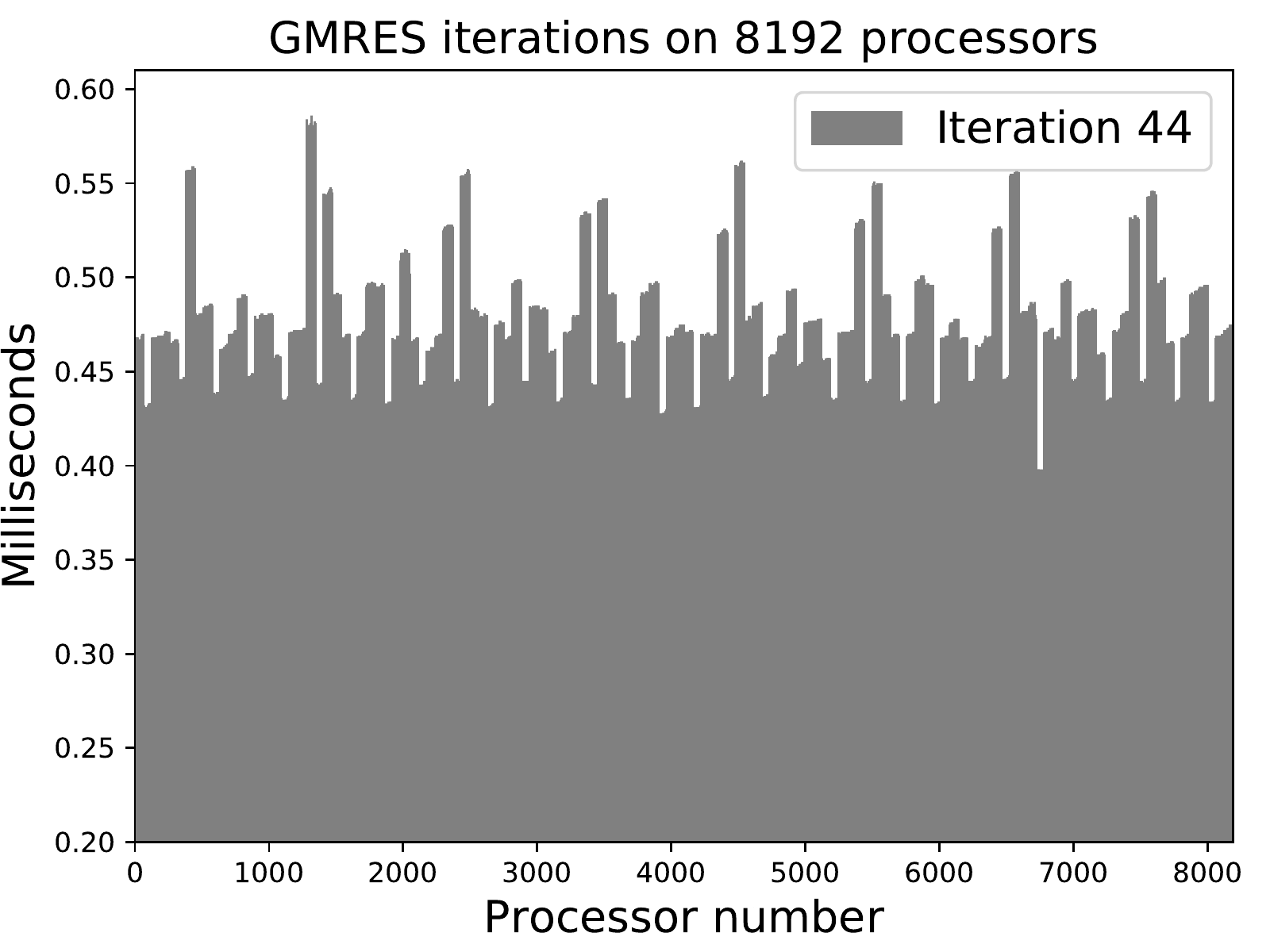}
\includegraphics[scale=0.5]{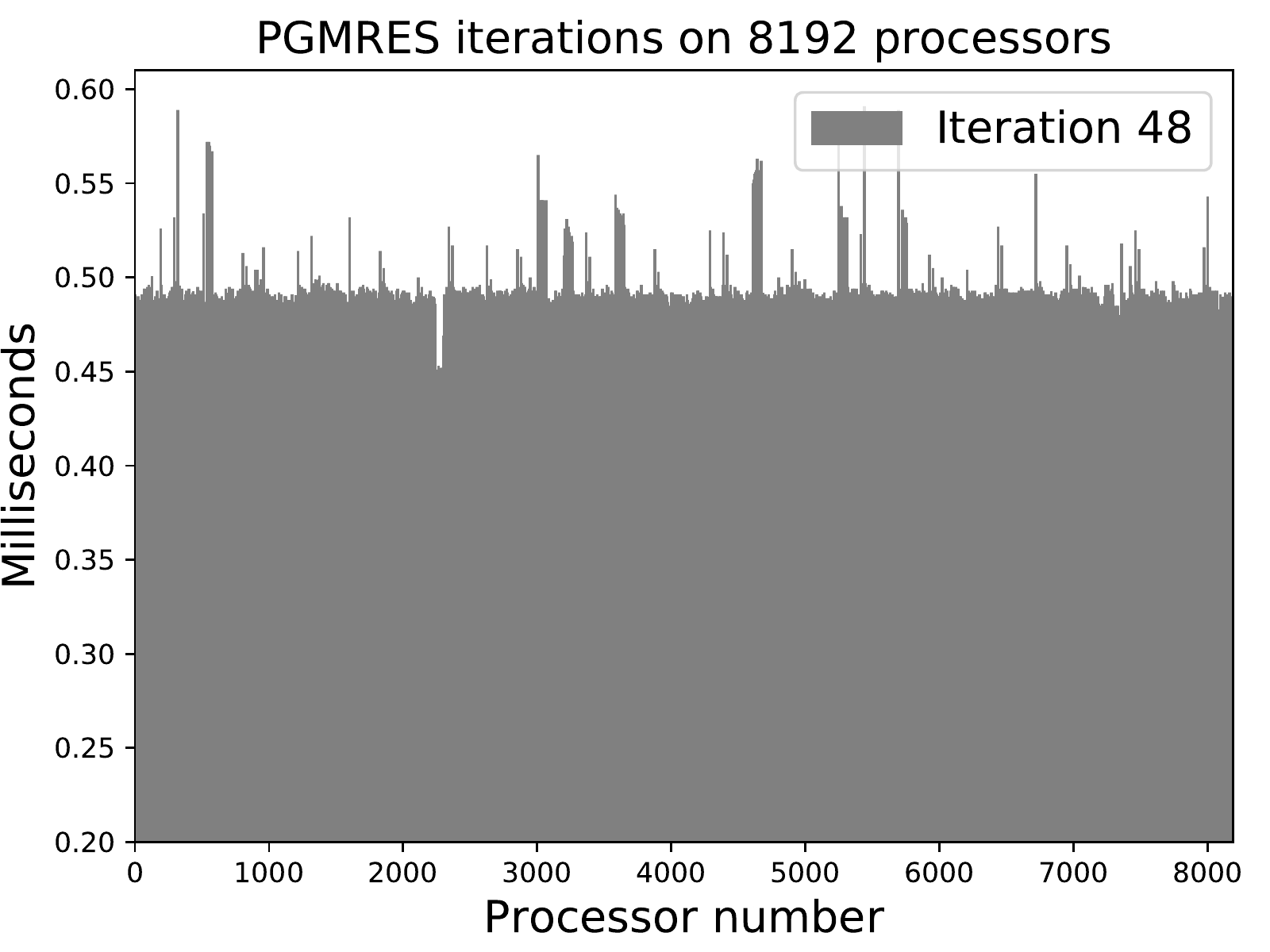} \\
\caption{GMRES (left) and PGMRES (right) iterations for PETSc ex48.} \label{fig:ex48}
\end{figure*}

PETSc SNES tutorial ex48\footnote{\texttt{src/snes/examples/tutorials/ex48.c} in PETSc 3.10} solves the Blatter-Pattyn hydrostatic equations that model ice sheet flow for glaciers (\cite{pattyn2008benchmark}).
For the ice, we use a power-law rheology with Glen exponent 3. This generates a much denser system of equations with about 10 times more nonzeros per row than ex23 and requires extensive communication across the machine, resulting in more network traffic and the possibility for more variable communication latency. Again we choose our problem size so that there are $10^6$ unknowns, use a Jacobi preconditioner, and force 5000 iterates of the {\color{blue} Krylov algorithm}. We stop after one nonlinear iteration.

We repeat the analysis from before and find qualitatively similar results, shown in Figure \ref{fig:ex48}. Using the Kolmogorov-Smirnov test again we find that we do not reject the assumption that GMRES iterates from ranks 0 and 1 come from the same distribution since $D = 0.003 < 0.024$ and we reject that PGMRES iterates come from the same distribution since $D = 0.078 > 0.023$.
The iterates on a fixed process $p$ look like they could be from the same family of distributions with different parameters as those in the previous subsection.

GMRES iterates again appear highly dependent on the node with node 0 variance $4.3\cdot10^{-13} \text{s}^2$. The variance on PGMRES node 0 is $8.6\cdot10^{-10} \text{s}^2$, larger than before, and there is more variation between PGMRES nodes which looks periodic. This could be explained by the communication induced by the matrix sparsity pattern.

\section{Non-stationary performance model} \label{sec:updated-model}

Based on the results from the last section, we move to a non-stationary performance model where the iterates $\mathcal{T}^k_p$ can fluctuate in time.
The expected total time for a Krylov method is still given by
$$E[T] = \sum_k E[ \max_p \mathcal{T}^k_p]$$
but we relax the claim that the iterates are stationary in time. In iteration $k$, the random variables $\mathcal{T}^k_p$ are drawn from a distribution with cdf $F_k(x)$ and pdf $f_k(x)$, which can now fluctuate across steps.
Then the expected runtime of a Krylov method is given by
\begin{equation}
\widehat{E}[T] =  \sum_k P \int ^{\infty}_{-\infty} x F_k(x)^{P-1} f_k(x) dx. \label{eq:updated-krylov-model},
\end{equation}
We call this the ``non-stationary" performance model and distinguish between $E[T]$ in \eqref{eq:krylov-model} and $\widehat{E}[T]$ here.

Similarly for a {\color{blue} pipelined algorithm}, we use \eqref{eq:pipeline-expression} as before and drop the assumption that the iterates are stationary in time, so that
\begin{equation}
\widehat{E}[T'] \rightarrow  \sum_k \mu_k \label{eq:updated-pipelined-model},
\end{equation}
where $\mu_k$ is the mean of the underlying distribution in iteration $k$.

The original Krylov model \eqref{eq:krylov-model} benefits from slight alteration as well.
Since the iterates on each node are nearly constant, shown in Figure \eqref{fig:ex23-independent} (left) and Figure \ref{fig:ex48} (bottom left), they are more suitably modeled by $\widehat{P} = P/64$  independent random variables than $P$ independent processors. We make use of this substitution in the next section but it does not seem to strongly affect \eqref{eq:updated-krylov-model}.

\section{Performance modeling} \label{sec:performance-model}

In this section, we will deploy our performance models and test them against collected data.

\subsection{Computing expected runtimes} \label{sec:computing-runtimes}

The expressions for $E[T]$ and $E[T']$ say that the iterates come from some underlying  distribution with pdf $f(x)$, cdf $F(x)$, and mean $\mu$.
We will use ``bulk statistics" with these performance models, taking all iterations for all processors in aggregate, for the underlying distribution and to characterize the iteration times and machine variation.

\begin{figure}[t]
\centering
\includegraphics[scale=0.5]{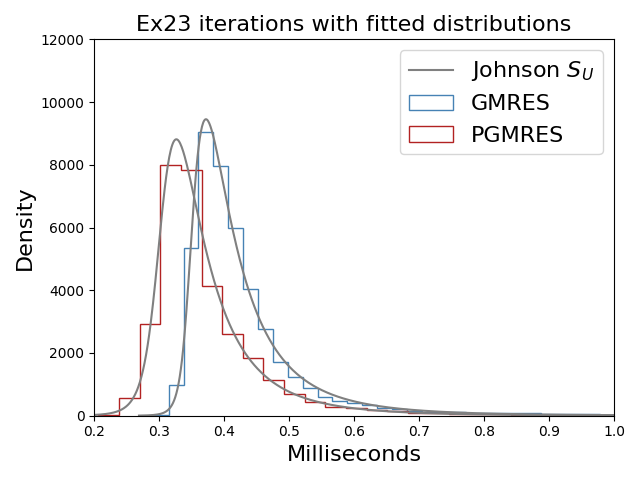}
\includegraphics[scale=0.5]{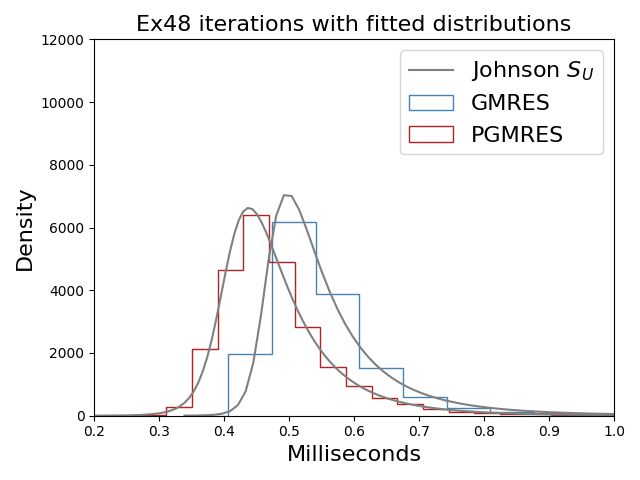}
\caption{GMRES and PGMRES bulk statistics with fitted distributions for PETSc ex23 (left) and ex48 (right).} \label{fig:bulk-fitted}
\end{figure}

\begin{figure}[b]
\centering
\includegraphics[scale=0.5]{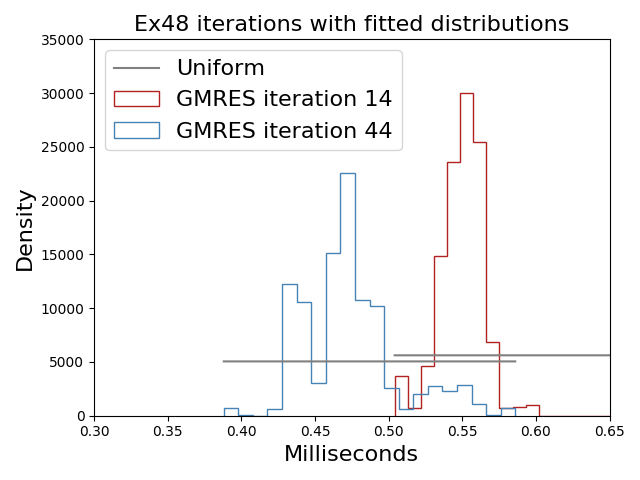}
\includegraphics[scale=0.5]{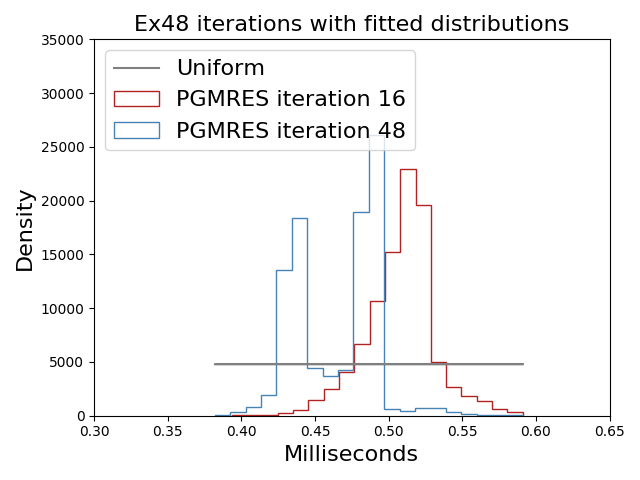}
\caption{GMRES  iterations $k = 14, 44$ and PGMRES $k = 16, 48$ iterations  with fitted uniform distributions.} \label{fig:iterations-with-fitted-uniform}
\end{figure}

To find the functions $f(x)$ and $F(x)$ and mean $\mu$, we use Scipy's ${\texttt{stats}}$ package to fit various analytical distributions to our collected data. For a given continuous distribution and data, the ${\texttt{fit}}$ function returns the maximum likelihood estimation for the shape, location, and scale parameters by maximizing a log-likelihood function.
We choose distributions which minimize the sum of squared error between the data and the fit distribution.
For the {\color{blue} pipelined algorithms}, we disregard the two iterations in each cycle that ``fill the pipeline" to avoid distribution fitting complications.

Using these well-fitting distributions, we can calculate $E[T]$ and $E[T']$ and compare the expected results to the {\color{blue} \texttt{KSPSolve} time} (the time spent inside the Krylov solver) provided by PETSc's ${\texttt{-log\_view}}$ option.
For integration in \eqref{eq:krylov-model} we use {\color{blue} Python's ${\texttt{scipy.integrate.quad}}$ function} and integrate over the bounds of the data.

To employ the updated models $\widehat{E}[T]$ and $\widehat{E}[T']$, we again perform distribution fitting using ${\texttt{scipy}}$, this time for each iteration to find  $f_k(x)$, $F_k(x)$, and mean $\mu_k$. {\color{blue}
GMRES and PGMRES iteration data is shown in Figure \ref{fig:iterations-with-fitted-uniform}.
It is quite clear that these are far from normal and do not resemble well-known distributions. Because of this, and in the interest of simplicity,
we treat runtimes $\mathcal{T}^k_p$ for each iteration $k$ as coming from a uniform distribution with possibly different parameters. }
{\color{blue} Again, integration} is performed in each iteration again using ${\texttt{scipy.integrate.quad}}$ and the calculated expected total times are compared to the \texttt{KSPSolve} time.

Analytical expressions have been used
 to bound the maximum of a set of random variables using the sample mean $\mu_X$ and standard deviation $\sigma_X$ (\cite{seelam2010extreme}).  Cramer (\cite{cramer2016mathematical, david2004order})
  bound the maximum of $N$ identical and independent random variables  with
\begin{equation}
E[X_{\max{(N)}}] \leq \mu_X + \frac{\sigma_X (N-1)}{\sqrt{2N-1}}
\end{equation}
and (Bertsimas \cite{bertsimas2006tight})
 bound the maximum of $N$ identical, but not independent random variables by
\begin{equation}
E[X_{\max{(N)}}] \leq \mu_X + \sigma_X \sqrt{N-1}.
\end{equation}
We will use these to compare with our Krylov performance models.

\subsection{Results}
Distribution fitting alone gives some insight into algorithm execution.
In Figure \ref{fig:bulk-fitted}, we show bulk data from our experimental results with the {\color{blue} Johnson $\text{S}_\text{U}$ distribution, a transformation of the normal distribution with pdf
\begin{equation}
f(x, a, b) = \frac{b}{\sqrt{x^2 + 1}} \phi(a + b \log(x + \sqrt{x^2 + 1}))
    \label{eq:johnsonsu_pdf}
\end{equation}
with shape paramters $a$ and $b$ and where $\phi$ is the normal pdf.}
Note that there were some other well-fitting distributions, including Non-central Student's T.
The bulk data show that in a given method, the iterates $\mathcal{T}^k_p$ are mostly clustered and fairly quick with a fewer, longer, giving the distribution a small tail.
In general, PGMRES runtimes were faster than GMRES, consistent with faster overall execution in these examples. Ex48 runtimes are shifted to the right of ex23 and more spread out, implying longer iterations with more variation.

\begin{table*}[t]
\small\sf\centering
\caption{Performance model results and bounds for ex23 and ex48 with good fitting distributions.} \label{tab:performance-model}
\begin{tabular*}{500pt}{@{\extracolsep\fill}l l l l l l l@{\extracolsep\fill}}
\toprule
\textbf{ex23} & \textbf{KSPSolve}  & \textbf{{\color{blue}$\boldsymbol{E_{\text {John}\text{S}_\text{U}}\left[T\right]}$}}  & \textbf{$\boldsymbol{E_{\text {NCT}}\left[T\right]}$}  & \textbf{$\boldsymbol{\widehat{E}_{\text {Unif}}\left[T\right]}$} & \textbf{Cramer bound}  & \textbf{Bertsimas bound} \\
\midrule
 GMRES & 2.217 & 4.831 & 4.365   & 2.432 & 6.35 & 8.102  \\
 PGMRES & 2.006 & 1.865 & 1.868 & 1.857   & & \\
\textbf{ex48} &  &  &  &  & & \\
 GMRES & 2.943 & 5.716 & 10.88  & 3.189 & 20.59  & 27.99 \\
 PGMRES & 2.656 & 2.413 & 2.413 & 2.455   & & \\
\bottomrule
\end{tabular*}
\end{table*}

\begin{table*}
\small\sf\centering
\caption{{\color{blue} Johnson $\text{S}_\text{U}$} parameters for uniform parameters $a_k$ and $s_k$.} \label{tab:distribution-params}
\begin{tabular*}{500pt}{@{\extracolsep\fill}lllllllll@{\extracolsep\fill}}
\toprule
&\multicolumn{4}{@{}c@{}}{\textbf{Uniform $a_k$}} & \multicolumn{4}{@{}c@{}}{\textbf{Uniform $s_k$}} \\\cmidrule{2-5}\cmidrule{6-9}
\textbf{ex23} & \textbf{${\texttt{$a$}}$}  & \textbf{${\texttt{$b$}}$} & \textbf{${\texttt{loc}}$}  & \textbf{${\texttt{scale}}$} & \textbf{${\texttt{$a$}}$}  & \textbf{${\texttt{$b$}}$} & \textbf{${\texttt{loc}}$}  & \textbf{${\texttt{scale}}$} \\
\midrule
 GMRES & -5.86e-01 & 3.35 & 3.97e-04 & 1.07e-21 & -7.40e-01 & 3.21 & 8.06e-04 & 1.86e-23 \\
 PGMRES & 2.84 & 6.74 & 3.10e-04 & 2.26e-19 & -6.18e-02 & 2.42 & 2.21e-03 & 6.47e-19 \\
 \textbf{ex48} &  &  &  &  & & & & \\
 GMRES & 6.74e-01 & 2.09 & 2.22e-02 & 2.24e-24 & 6.69e-01 & 2.10 & 1.71e-03 & 9.23e-26 \\
 PGMRES & -6.02e-01 & 3.34 & 4.07e-04 & 3.05e-23 & 1.24 & 1.84 & 1.36e-02 & 1.66e-18 \\
\bottomrule
\end{tabular*}
\end{table*}

Results for the data presented in our experimental results are shown in Table \ref{tab:performance-model} for PETSc ex23 and ex48.
{\color{blue}
It is clear that applying the non-stationary performance model $\widehat{E}[T]$ to standard Krylov subspace methods gives a large improvement over the stationary model $E[T]$. The stationary model  overestimates actual runtimes and the analytical bounds are very loose and not particularly helpful.
Both models are good estimates for {\color{blue} pipelined algorithms}, suggesting that modeling a method without explicit synchronization is more flexible.

In this work we don't compare our results to more algorithm-specific models or those that collect detailed hardware measurements, instead emphasizing a paradigm where the use of simple, general models is sufficient to approximate the execution time for Krylov solvers without collecting low-level information.}

\section{Predicting runtimes}\label{sec:performance-estimates}

In the previous section, we showed that our performance models $\widehat{E}[T]$ and $\widehat{E}[T']$ can reasonably predict execution time when we have the time for iteration $k$ on process $p$ for all $k$ and $p$.

\begin{figure}[b]
\centering
\includegraphics[scale=0.5]{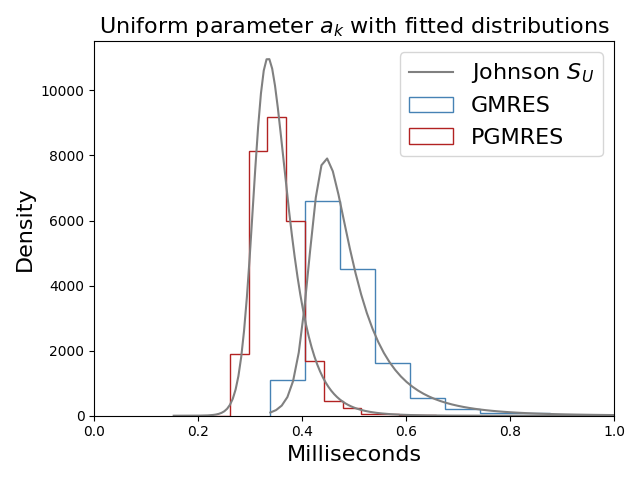}
\includegraphics[scale=0.5]{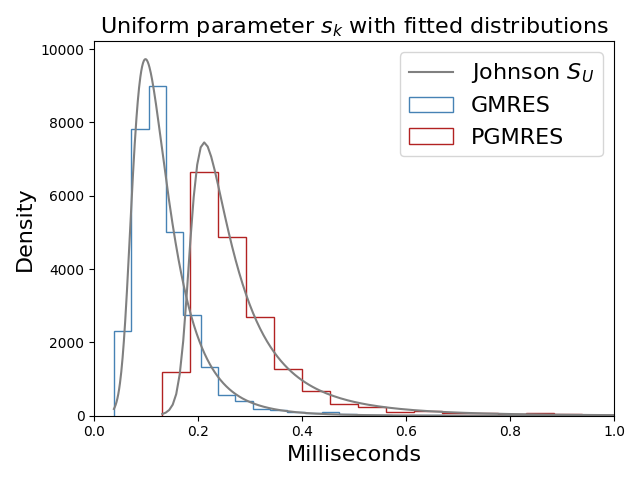}
\caption{Uniform parameters $a_k$ and $s_k$ from PETSc ex48 with fitted distributions.} \label{fig:uniform-params-fitted}
\end{figure}

{\color{blue} While bulk} statistics appear to be enough to succinctly describe the performance of a pipelined {\color{blue} algorithm}, they are not a sufficient way to describe a traditional Krylov method. {\color{blue} Instead, we will use non-stationary  distributions to model each iteration.
GMRES and PGMRES distribution data is shown with uniform distributions in Figure \ref{fig:iterations-with-fitted-uniform}.
Here we see how much the iterations shift in time. They are mostly non-overlapping.
We also want to know how the uniform parameters change over time. }
That is, for each iteration $k$, the random variables $\mathcal{T}_p^k$ are modeled as following a uniform distribution with pdf $f_k$ and cdf $F_k$ given by
\begin{equation} \label{eq:uniform}
f_k(x) = \frac{1}{b_k - a_k}, \quad F_k(x) = \frac{x-a_k}{b_k-a_k}.
\end{equation}
The uniform distribution can be described with two parameters: the minimum $a_k$ and the span $s_k = b_k - a_k$. We model a Krylov method with $K$ uniform distributions
where in each iteration the uniform parameters $a_k$ and $s_k$ are random variables themselves drawn from some distribution.
Figure \ref{fig:uniform-params-fitted} show histograms of the uniform parameters from  PETSc ex48 using GMRES and PGMRES with fitted {\color{blue} Johnson $\text{S}_\text{U}$} distribution.
This distribution {\color{blue} has the pdf \eqref{eq:johnsonsu_pdf} and} is described by two shape parameters $a$ and $b$ and location and scale parameters ${\texttt{loc}}$ and ${\texttt{scale}}$ in Scipy .

Table \ref{tab:distribution-params} shows {\color{blue} Johnson $\text{S}_\text{U}$} parameters for the  uniform distribution parameters from Figure \ref{fig:uniform-params-fitted} and PETSc ex23. In general, they show that, in a given iteration, the fastest processor is generally faster in a PGMRES run, but the runtimes are more spread out. All {\color{blue} Johnson $\text{S}_\text{U}$} distributions are shaped such that they lean left with a tail, some longer than others.

{\color{blue} We can mimic a Krylov computation by ``simulating'' the iterations with a Monte Carlo-type approach.
We assume} that each iteration time $\mathcal{T}_p^k$ is uniformly distributed with parameters $a_k$ and $s_k$ in iteration $k$
$$\mathcal{T}_p^k \sim \text{Uniform}(a_k, s_k).$$
We further assume that the parameters $a_k$ and $s_k$ are themselves random variables drawn from a {\color{blue} Johnson $\text{S}_\text{U}$} distribution with parameters $a$, $b$, $\texttt{loc}$, and $\texttt{scale}$.

We draw random variables $a_k$ and $s_k$  using Scipy's $\texttt{rvs}$ {\color{blue} function and compute} the expected time for iteration $k$, $\widehat{E}[T_k]$, given by
\begin{equation}
\widehat{E}[T_k] =  \widehat{P} \int ^{\infty}_{-\infty} x F_k(x)^{\widehat{P}-1} f_k(x) dx.
\end{equation}
Repeating for $K$ iterations, we get $\widehat{E}[T]$.
We simulate a pipelined Krylov computation in the same way, pulling random variables $a_k$ and $s_k$ and computing
\begin{equation}
\widehat{E}[T'_k] =  \mu_k
\end{equation}
where $\mu_k$ is the mean of $\text{Uniform}(a_k, s_k)$.
Table \ref{tab:simulated-runtimes} shows the results of this simulation using the {\color{blue} Johnson $\text{S}_\text{U}$} parameters from Table \ref{tab:distribution-params}. The results are good when we have the computed {\color{blue} Johnson $\text{S}_\text{U}$} parameters. The challenge in general will be to reason about these parameters so that we can make a priori performance estimates for {\color{blue} standard Krylov methods and pipelined algorithms}.

\begin{table*}
\small\sf\centering
\caption{Actual and simulated runtimes in seconds on 8192 processors with $10^6$ unknowns.} \label{tab:simulated-runtimes}
\begin{tabular*}{500pt}{@{\extracolsep\fill}llllll@{\extracolsep\fill}}
\toprule
\textbf{ex23} & \textbf{GMRES}  & \textbf{PGMRES}  & \textbf{ex48}  & \textbf{GMRES} & \textbf{PGMRES} \\
\midrule
 KSPSolve & 2.217 & 2.006 &  KSPSolve & 2.943 & 2.656 \\
 Simulated & 2.416 & 1.908 &  Simulated &  3.14 & 2.482  \\
\bottomrule
\end{tabular*}
\end{table*}

\section{Extension to other algorithms and computing platforms}\label{sec:more-experiments}

In the last section, {\color{blue} we saw that we can reasonably estimate the execution time of a Krylov method given that we know the {\color{blue} Johnson $\text{S}_\text{U}$} parameters that model $a_k$ and $s_k$.
Many factors, such as algorithm and matrix pattern, can influence performance.
In this section, we expand our experiments to study other factors including problem size and computing platform and test our non-stationary model in more scenarios.}

\subsection{Scaling experiments}\label{sec:scaling}

\begin{figure}[t]
\centering
\includegraphics[scale=0.5]{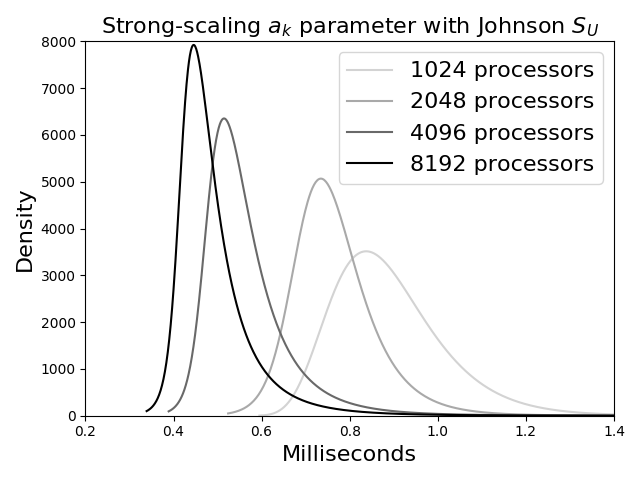}
\includegraphics[scale=0.5]{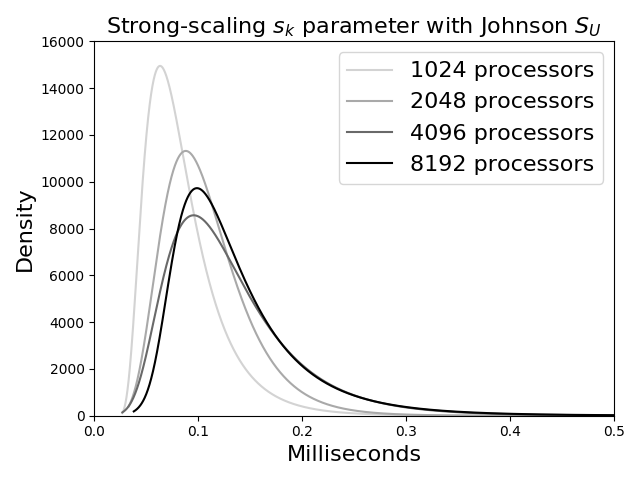}
\caption{GMRES ex48 strong-scaling results on $P = 1024 - 8192$ processors.} \label{fig:strong-scaling}
\end{figure}

\begin{figure}[b]
\centering
\includegraphics[scale=0.5]{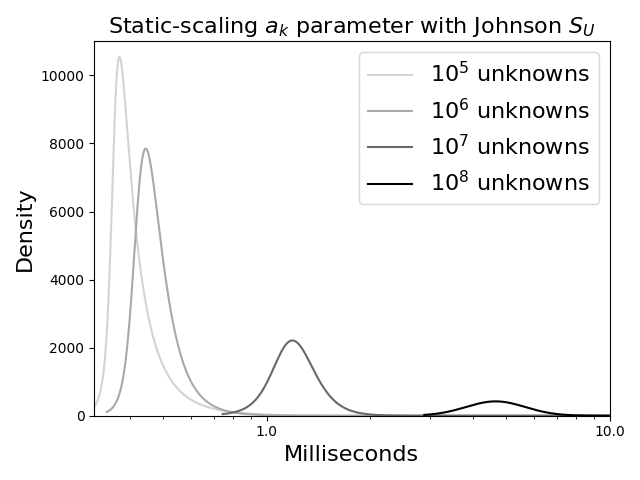}
\includegraphics[scale=0.5]{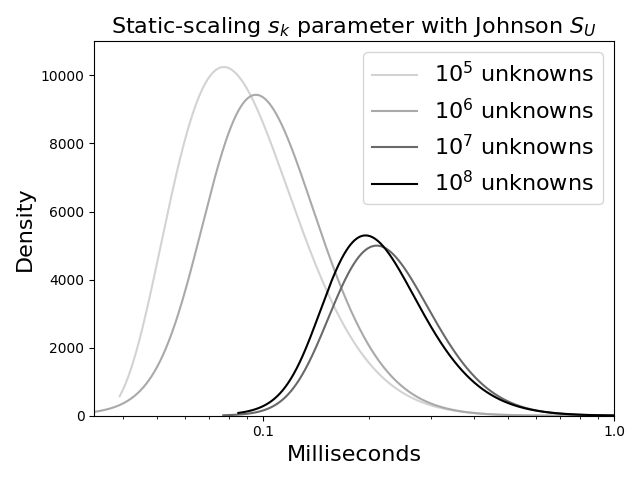}
\caption{GMRES ex48 static-scaling results for $10^5 - 10^8$ unknowns.} \label{fig:static-scaling}
\end{figure}

Our experiments so far have been performed on 8192 processors with $10^6$ unknowns.
To see how changing processor count and problem size affect the Krylov method performance and underlying uniform parameters, we run strong- and static-scaling experiments.
Figures \ref{fig:strong-scaling} and \ref{fig:static-scaling} show the {\color{blue} Johnson $\text{S}_\text{U}$} distributions for the uniform parameters $a_k$ and $s_k$ for strong- and static-scaling experiments on Theta.

With strong-scaling experiments, we see how varying the processor count affects Krylov method performance for a fixed problem. We repeat runs of ex48 with  $10^6$ unknowns on $P = 1024 - 8192$ processors.
We see that, for a fixed number of unknowns, as we decrease the number of processors the {\color{blue} Johnson $\text{S}_\text{U}$} distribution shifts to the right and spreads out for the uniform $a_k$ parameter.
In this case, each processor does more work and is more likely to experience a detour.
Similarly, we perform static-scaling experiments.
In static-scaling, the problem size (or number of unknowns) is varied on a fixed number of processors.
This allows us to see the {\color{blue} effect of computation without changing} the communication pattern. We keep the number of processors fixed at $P=8192$ and vary the number of unknowns, which exaggerates what we see for strong-scaling as shown in Figure \ref{fig:static-scaling}.


\subsection{BiCGSTAB}\label{sec:bcgs}

We solve PETSc ex23, the one-dimensional Laplacian problem, using the Stabilized Biconjugate Gradient method (BiCGSTAB) (\cite{van1992bi})
and a pipelined version (PIPEBiCGSTAB) (\cite{cools2017communication})
with 5000 linear iterations and $10^6$ unknowns on Theta.
We choose these methods because their communication and performance characteristics differ somewhat from GMRES, as the methods are based on a short-term recurrence relation and do not require orthogonalizing against a growing set of basis vectors; BiCGSTAB and variants also require two matrix-vector multiplications per iterations, instead of one.

\begin{figure}[t]
\centering
\includegraphics[scale=0.5]{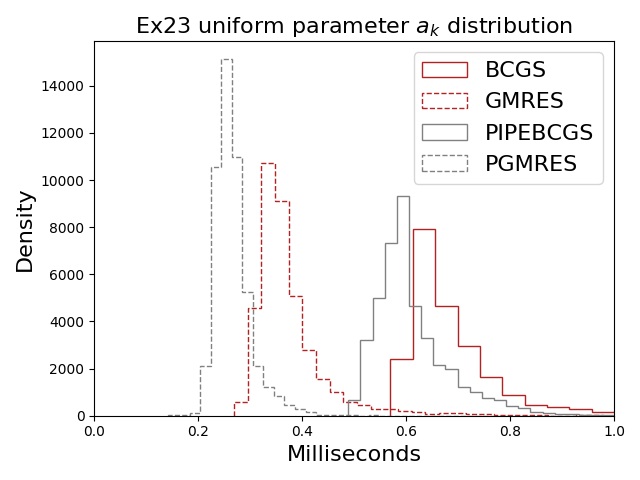}
\includegraphics[scale=0.5]{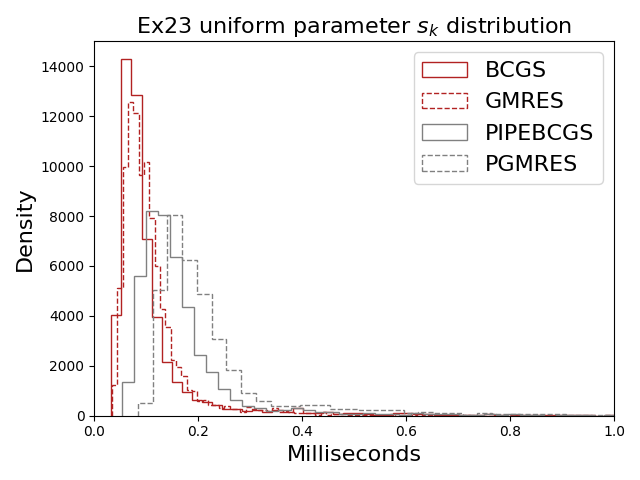}
\caption{BiCGSTAB and PIPEBiCGSTAB uniform parameters compared to GMRES and PGMRES.} \label{fig:bcgs}
\end{figure}

Many aspects of the BiCGSTAB and PIPEBiCGSTAB computations are consistent with GMRES and PGMRES, such as non-stationary iterates and nearly constant BiCGSTAB runtimes on a given node of Theta.
Figure \ref{fig:bcgs} shows uniform parameter $a_k$ and $s_k$ histograms with GMRES and PGMRES for comparison.
{\color{blue} It appears that BiCGSTAB and PIPEBiCGSTAB distributions can be modeled by the {\color{blue} Johnson $\text{S}_\text{U}$} distribution family. This
suggests} that our performance model is applicable to Krylov methods outside of GMRES methods.
GMRES and PGMRES outperform BiCGSTAB and PIPEBiCGSTAB and have quicker minimum iteration times ({\color{blue} Johnson $\text{S}_\text{U}$} distributions shifted to the left for uniform parameter $a_k$), but runtimes are  tightly grouped.
Performance results for BiCGSTAB and PIPEBiCGSTAB using our non-stationary stochastic models are in shown Table \ref{tab:performance-model-2}.

\subsection{Piz Daint}\label{sec:pizdaint}

\begin{figure}[t]
\centering
\includegraphics[scale=0.5]{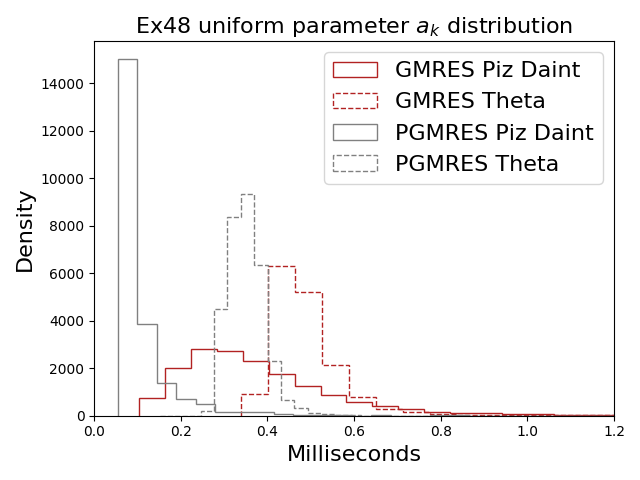}
\includegraphics[scale=0.5]{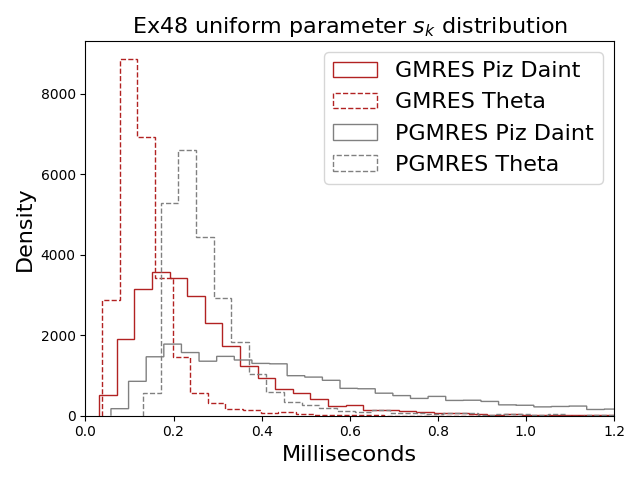}
\caption{GMRES and PGMRES on Theta and Piz Daint.} \label{fig:pizdaint}
\end{figure}

We repeat experiments on the Cray XC40 Piz Daint supercomputer\footnote{hybrid partition, November 2018, PETSc 3.10, Cray-MPICH 7.7.2} at the Swiss National Computing Center.
Piz Daint and Theta both employ a Aries Dragonfly interconnect but Piz Daint contains Intel Xeon E5 Haswell  processors (\cite{hammarlund2014haswell})
on compute nodes {\color{blue} (which provide more performance per CPU core compared to KNL). Therefore, the differences we see} are likely due to different processors or the shared use of a communication network.
Figure \ref{fig:pizdaint} shows runs of ex48 with GMRES and PGMRES on 8192 processors and $10^6$ unknowns.
In a given iteration, the quickest processor can be much faster on Piz Daint than Theta, but both uniform parameters contain much more variation.
Non-stationary performance model results from Piz Daint are in shown Table \ref{tab:performance-model-2}.

\begin{table*}
\small\sf\centering
\caption{Non-stationary performance model results for ex23 and ex48 for extended experiments.} \label{tab:performance-model-2}
\begin{tabular*}{500pt}{@{\extracolsep\fill}llllllll@{\extracolsep\fill}}
\toprule
&\multicolumn{2}{@{}c@{}}{\textbf{Theta}} &  & \multicolumn{2}{@{}c@{}}{\textbf{Piz Daint}} & \multicolumn{2}{@{}c@{}}{\textbf{Mira}} \\\cmidrule{2-3}\cmidrule{5-6} \cmidrule{7-8}
\textbf{ex23} & \textbf{BiCGSTAB}  & \textbf{PIPEBiCGSTAB}  &\textbf{ex48} & \textbf{GMRES}  & \textbf{PGMRES} & \textbf{GMRES}  & \textbf{PGMRES}    \\
\midrule
 KSPSolve & 3.957 & 3.53 & KSPSolve & 3.029 & 2.642 & 3.026 & 2.804 \\
 $\widehat{E}_{\text {Unif}}\left[T\right]$ & 4.227 & 3.512 & $\widehat{E}_{\text {Unif}}\left[T\right]$ & 3.541 & 2.473 & 3.189 & 2.455 \\
\bottomrule
\end{tabular*}
\end{table*}

\begin{figure}[b]
\centering
\includegraphics[scale=0.5]{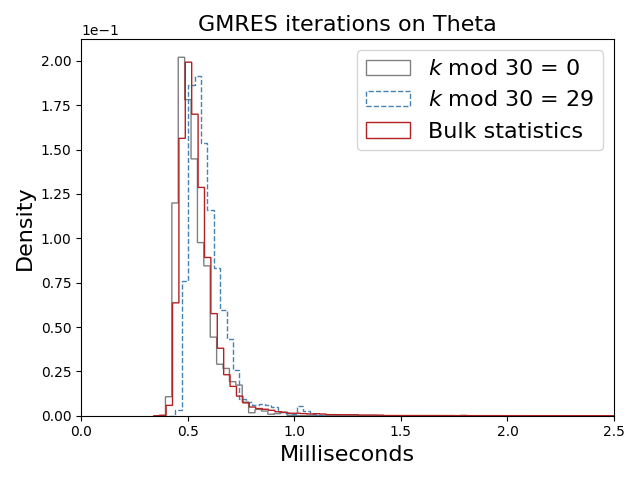}
\includegraphics[scale=0.5]{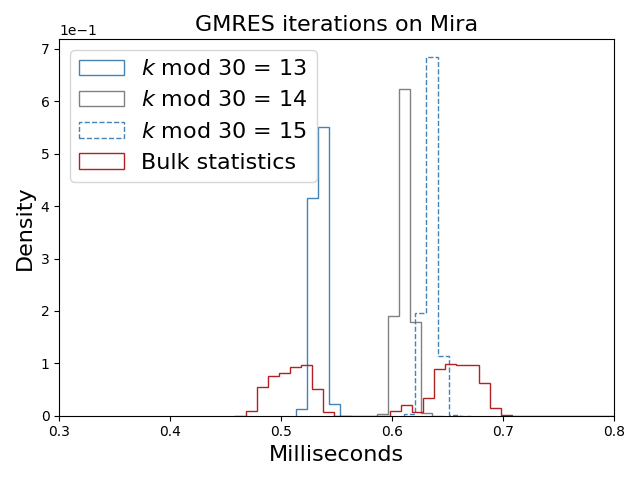}
\caption{GMRES ex48 iteration density and Theta and Mira.}  \label{fig:mira}
\end{figure}

\subsection{Mira}\label{sec:mira}

We repeat ex48 experiments on Mira, an IBM Blue Gene/Q\footnote{November 2018, PETSc 3.10, MPICH2 1.5} (\cite{kumaran2016introduction}) at Argonne's ALCF.
We expect a marked performance difference on Mira since
nodes on Mira are connected by a 5D torus network with special hardware support for collective communication. Conversely, nodes on Theta are connected by a Cray Aries Network where neighboring workloads share network routes, increasing performance variability (\cite{groves2017understanding}).
Each Blue Gene/Q node also contains a redundant processor to service operating system interrupts, similar to Cray's core specialization feature where one core per node is dedicated to handling OS operations, which reduces core variability (\cite{chunduri2017run}).

Figure \ref{fig:mira} shows histograms of iterations from difference places in the GMRES restart cycle as well as a bulk histogram for all iterations from all processors, which can be thought of as the average performance. Each curve has been normalized to represent a distribution.
GMRES iterations on Theta at different places in the cycle appear similar to the average. They are noisy enough that our simulation provided reasonable results even though we drew uniform parameters $a_k$ and $s_k$ from the same {\color{blue} Johnson $\text{S}_\text{U}$} distributions regardless of $k$ in the GMRES cycle.
With Mira's quiet network, the GMRES iterates are much more distinct suggesting that a refined model that accounts for the amount of work done per iteration would be needed for a priori estimates.
Futhermore, we see a jump in the time between iterations $k \mod 30 = 13$ and $k \mod 30 = 14$, which is mostly likely due to L1 data cache effects as the required storage for basis vectors increases throughout a GMRES cycle.
Similar results were found for PGMRES and performance results are in Table \ref{tab:performance-model-2}.

\section{Conclusion and future work} \label{sec:conclusion}
In this work we gather fine-grained iteration data from runs of various {\color{blue} Krylov subspace methods} across multiple HPC platforms and for different computations.
We examine the iterates over time and find that they are non-stationary. {\color{blue} That is, the distribution of runtimes shifts with time even after accounting for the growing Krylov subspace.}
{\color{blue} We see significant inter-node variability as well as some intra-node variability, possibly from reduction operations, synchronizations, or operating system interrupts, hindering the performance of traditional
Krylov methods.}
Inhomogeneity from communication patterns is another source of variation between processor performance.

Using these insights, we develop a non-stationary, stochastic performance model
that accurately reflects the performance of {\color{blue} standard Krylov methods and pipelined algorithms}
across different algorithms and a variety of HPC architectures.
Furthermore, this model suggests a way to make predictive performance estimates.

{\color{blue} Future work should conduct experiments on a more comprehensive set of problems, gathering a wider range of runtimes for testing. We also plan to measure different components of Krylov iterations such as communication phases with orthogonalizations and make use of hardware counters.}
Also in the future, nondeterministic performance models can be derived for other parallel algorithms
where unpredictable system interference could impact performance.
We also plan on testing our model with heavier preconditioners, deeper pipelines, and more complex architectures. Since
we made no assumptions about the hardware in our performance models, we could deploy the same experiments on,
for example, a heterogenous machine with GPUs and analyze the data in a similar way.

{\color{blue} Accurate performance models can assist in} algorithmic choices and parameter tuning such as pipeline depth.
Insights from performance models can also be used to guide the development of new algorithms,
particularly those we expect to be running in less predictable computing environments,
such as heavily loaded machines or loosely coupled networks used in cloud computing
or those with shared resources.

\section*{Acknowledgements}
Thanks to everyone who offered insight and advice, including Todd Munson, Barry Smith, Ivana Marincic, Vivak Patel, Karl Rupp, and Oana Marin.

\section*{Funding}
The authors were supported by the U.S. Department of Energy, Office of Science, Advanced Scientific Computing Research un- der Contract DE-AC02-06CH11357. This research used resources of the Argonne Leadership Computing Facility, which is a DOE Office of Science User Facility supported under Contract DE-AC02- 06CH11357. The authors were partially supported by the Exascale Computing Project (17-SC-20-SC), a collaborative effort of the U.S. Department of Energy Office of Science and the National Nuclear Security Administration. P.~S. acknowledges financial support from the Swiss University Conference and the Swiss Council of Federal Institutes of Technology through the Platform for Advanced Scientific Computing (PASC) program.

\bibliographystyle{unsrt}
\bibliography{bibli}

\clearpage
\onecolumn
\center
\framebox{
   \parbox{4in}{
   The submitted manuscript has been created by UChicago Argonne, LLC, Operator of Argonne
   National Laboratory (``Argonne’’). Argonne, a U.S. Department of Energy Office of Science
   laboratory, is operated under Contract No. DE-AC02-06CH11357. The U.S. Government retains
   for itself, and others acting on its behalf, a paid-up nonexclusive, irrevocable worldwide
   license in said article to reproduce, prepare derivative works, distribute copies to the
   public, and perform publicly and display publicly, by or on behalf of the Government.
   The Department of Energy will provide public access to these results of federally
   sponsored research in accordance with the DOE Public Access Plan.
   \url{http://energy.gov/downloads/doe-public-accessplan}
   }
}

\end{document}